\documentclass[prl,twocolumn,showpacs,superscriptaddress]{revtex4}
\usepackage{graphicx}
\usepackage[tbtags]{amsmath}

\newcommand{\pht}{{\vphantom{2}}}
\newcommand{\ij}{i\kern -0.08em j}

\begin{document}

\title{Continuous Monitoring of Rabi Oscillations in a Josephson Flux Qubit}

\author{E. Il'ichev}
\affiliation{Institute for Physical High Technology, P.O. Box 100239, D-07702 Jena, Germany}
\author{N.~Oukhanski}
\affiliation{Institute for Physical High Technology, P.O. Box 100239, D-07702 Jena, Germany}
\author{A.~Izmalkov}
\affiliation{Institute for Physical High Technology, P.O. Box 100239, D-07702 Jena, Germany}
\affiliation{Moscow Engineering Physics Institute (State University), Kashirskoe sh.\ 31, 115409 Moscow, Russia}
\author{Th.~Wagner}
\affiliation{Institute for Physical High Technology, P.O. Box 100239, D-07702 Jena, Germany}
\author{M.~Grajcar}
\affiliation{Department of Solid State Physics, Comenius University, SK-84248 Bratislava, Slovakia}
\affiliation{Friedrich Schiller University, Institute of Solid State Physics, D-07743 Jena, Germany}
\author{H.-G. Meyer}
\affiliation{Institute for Physical High Technology, P.O. Box 100239, D-07702 Jena, Germany}
\author{A.Yu.\ Smirnov}
\affiliation{D-Wave Systems Inc., 320-1985 W. Broadway, Vancouver, B.C., V6J 4Y3, Canada}
\author{Alec Maassen van den Brink}
\affiliation{D-Wave Systems Inc., 320-1985 W. Broadway, Vancouver, B.C., V6J 4Y3, Canada}
\author{M.H.S. Amin}
\affiliation{D-Wave Systems Inc., 320-1985 W. Broadway, Vancouver, B.C., V6J 4Y3, Canada}
\author{A.M. Zagoskin}
\email{zagoskin@dwavesys.com}
\affiliation{D-Wave Systems Inc., 320-1985 W. Broadway, Vancouver, B.C., V6J 4Y3, Canada}
\affiliation{Physics and Astronomy Dept., The University of British Columbia, 6224 Agricultural Rd., Vancouver, B.C., V6T 1Z1, Canada}

\date{\today}

\pacs{03.67.Lx
, 85.25.Cp
, 85.25.Dq}

\begin{abstract}
Under resonant irradiation, a quantum system can undergo coherent (Rabi) oscillations in time. We report evidence for such oscillations in a \emph{continuously} observed three-Josephson-junction flux qubit, coupled to a high-quality tank circuit tuned to the Rabi frequency. In addition to simplicity, this method of \emph{Rabi spectroscopy} enabled a long coherence time of about 2.5$\mu$s, corresponding to an effective qubit quality factor $\sim$7000.
\end{abstract}

\maketitle

Small superconducting devices can exist in superpositions of macroscopically distinct states, making them potential qubits~\cite{mss}. In \emph{flux qubits} such states are distinguished by a magnetic flux, corresponding to a circulating current $\lesssim1\mu$A. Microwaves in resonance with the spacing between a qubit's energy levels will cause their occupation probabilities to oscillate, with a frequency proportional to the microwave amplitude. Such \emph{Rabi oscillations} have been detected, using statistical analysis of the response to external pulses \cite{NakamuraPashkinTsai,Esteve,Martinis,NakamuraDelft}. This requires precise pulse timing and shape, motivating our search for a complementary, continuous observation method.

The oscillations will only last
for a coherence time from the instant the high-frequency
(HF) signal is applied, after which the levels will be occupied
almost equally. The system's \emph{correlations}, however, will be
affected by Rabi oscillation as long as the HF signal is on.
Therefore, the signature of Rabi oscillations can be found in the autocorrelation function or its Fourier transform, the spectral density. In particular, if the
qubit is coupled to a tank circuit, the spectral density of the
tank-voltage fluctuations rises above the background noise when
the qubit's Rabi frequency~$\omega_\mathrm{R}$ coincides with
the tank's resonant frequency~$\omega_\mathrm{T}$. This forms
the basis for our measurement procedure of \emph{Rabi spectroscopy}.

The qubit is described by the Hamiltonian
\begin{equation}\label{eq_Ham}
  H=-{\textstyle\frac{1}{2}}(\Delta\sigma_{\!x} + \epsilon\sigma_{\!z}) - W(t) \sigma_{\!z}\;,
\end{equation}
with $\Delta$ the tunnel splitting, $\epsilon$ the DC
energy bias, and $W(t)= W \cos \omega_{_{\rm HF}}t$
the HF signal. For $W=0$, the qubit levels $\left|0\right>$ and
$\left|1\right>$ have energies $\mp\frac{1}{2}\Omega$
respectively, where $\Omega = \sqrt{\Delta^2+\epsilon^2}$. Rabi
oscillations between these levels are induced near resonance
$\omega_{_{\rm HF}}\approx \Omega/\hbar$. \emph{At} resonance,
\begin{equation}\label{eq_Rabi_freq}
   \hbar\omega_\mathrm{R}= (\Delta/\Omega)\, W\;,
\end{equation}
proportional to $W$ ($\propto\sqrt{P}$, with $P$ the HF power)~\cite{prob}.

Changes in qubit magnetic moment due to Rabi oscillation excite the tank. If photons in the latter have a sufficiently long lifetime, one can accumulate (hence the name ``tank") a sufficient number for their detection to be almost classical, while (for a small coupling coefficient) their emission from the qubit is still weak enough to disrupt the Rabi oscillation only occasionally. In detail, the qubit contribution to the nonequilibrium spectral density of the tank voltage is~\cite{footnote}
\begin{align}\label{S}
 S_V(\omega,\omega_\mathrm{R})&=2\frac{\epsilon^2}{\Omega^2}
 k^2\frac{L_\mathrm{q}^\pht I_\mathrm{q}^2}{C_\mathrm{T}}\omega^2\Gamma(\omega)
 \frac{\omega_\mathrm{R}^2}
      {(\omega^2{-}\omega_\mathrm{R}^2)^2+\omega^2\Gamma^2(\omega)}\notag\\
 &\quad\times\frac{\omega_\mathrm{T}^2}
      {(\omega^2{-}\omega_\mathrm{T}^2)^2+\omega^2\gamma_\mathrm{T}^2}\;,
\end{align}
where $\gamma_\mathrm{T}$ ($C_\mathrm{T}$) is the damping rate (capacitance) of the tank circuit, and $L_\mathrm{q}$ ($I_\mathrm{q}$) is the inductance (persistent current) of the qubit loop. The tank--qubit coupling is $k\equiv M/\sqrt{L_\mathrm{T}L_\mathrm{q}}$, with $M$ ($L_\mathrm{T}$) the mutual (tank) inductance. The effect vanishes at the degeneracy point $\epsilon=0$, since both qubit eigenstates then have zero average persistent current~\cite{VdW}. Note that $S_V$ in Eq.~(\ref{S}) has two factors: the first describes the qubit signal, the second its filtering by the tank's response function. Besides, the total tank spectrum $S_{V,\mathrm{t}}(\omega,\omega_\mathrm{R})$ has a background part $S_\mathrm{b}(\omega)$, due to thermal and quantum damped-oscillator noise uncorrelated with the qubit contribution.

The qubit's total decoherence rate,
\begin{equation}\label{Gtot}
 \Gamma(\omega)=\Gamma_0+4k^2\frac{L_\mathrm{q}^\pht I_\mathrm{q}^2}{\hbar^2}
 \frac{\epsilon^2}{\Omega^2}\omega_\mathrm{T}^2
 \frac{\gamma_\mathrm{T}^\pht k_\mathrm{B}T}
   {{(\omega^2{-}\omega_\mathrm{R}^2)}^2+\gamma_\mathrm{T}^2\omega_\mathrm{R}^2}\;,
\end{equation}
incorporates an internal contribution $\Gamma_0$, as well as a resonant one of the tank; see further Ref.~\cite{footnote}. If $\Gamma=\Gamma(\omega_\mathrm{T})\gg\gamma_\mathrm{T}$, the peak value in $\omega$ of $S_V$ is reached at~$\omega_\mathrm{T}$:
  $S_{V,\mathrm{max}}(\omega_\mathrm{R})\propto\omega_\mathrm{R}^2/
  [(\omega_\mathrm{T}^2{-}\omega_\mathrm{R}^2)^2+\omega_\mathrm{T}^2\Gamma^2]$.
Thus, $S_{V,\mathrm{max}}(\omega_\mathrm{R})$ lies on a Lorentzian-type curve with a width determined by~$\Gamma$. Hence, observing such a distribution not only is evidence for Rabi oscillations in the qubit, but also gives information about its coherence time.

\begin{figure}
\includegraphics[height=3in,angle=-90]{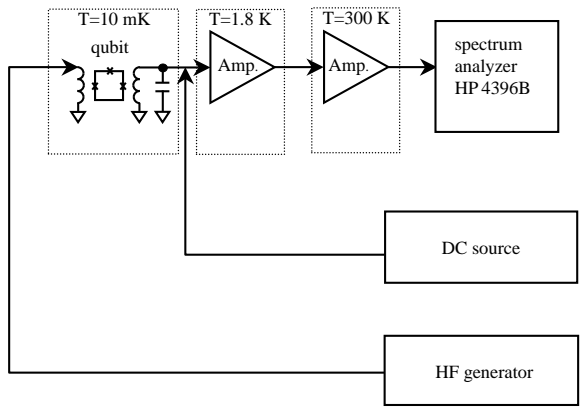}
\caption{Measurement setup. The flux qubit is inductively coupled
to a tank circuit. The DC source applies a constant flux
$\Phi_\mathrm{e}\approx\frac{1}{2}\Phi_{0}$. The HF generator
drives the qubit through a separate coil at a frequency close to
the level separation $\Delta/h=868$MHz. The output voltage at the
resonant frequency of the tank is measured as a function of HF
power.}\label{fig1}
\end{figure}

We use a small-inductance superconducting loop interrupted by
three Josephson junctions (a 3JJ qubit)~\cite{Mooij1}, inductively
coupled to a high-quality superconducting tank
circuit~\cite{Ilichev01} (Fig.~\ref{fig1}). This approach is similar to the one in entanglement experiments with Rydberg atoms and microwave photons in a cavity~\cite{Raimond01}. The tank serves as a sensitive
detector of Rabi transitions in the qubit, and simultaneously as a
filter protecting it from noise in the external circuit. Since
$\omega_\mathrm{T}\ll\Omega/\hbar$, the qubit is effectively
decoupled from the tank unless it oscillates with
frequency~$\omega_\mathrm{T}$. That is, while \emph{wide}-band (i.e., fast on the qubit time scale) detectors up to now have received most theoretical attention (e.g.,~\cite{KA}), we use \emph{narrow}-band detection to have sufficient sensitivity at a single frequency even with a small coupling coefficient; cf.\ above Eq.~(\ref{S}). The tank voltage is amplified and
sent to a spectrum analyzer. This is a development of the
Silver--Zimmerman setup in the first RF-SQUID
magnetometers~\cite{Zimm}, and is effective for probing flux
qubits \cite{Greenberg02a,Greenberg02b}. As such, it was used to
determine the potential profile of a 3JJ qubit in the classical
regime~\cite{Il'ichev-APL}.

\begin{figure}
\includegraphics[width=3in]{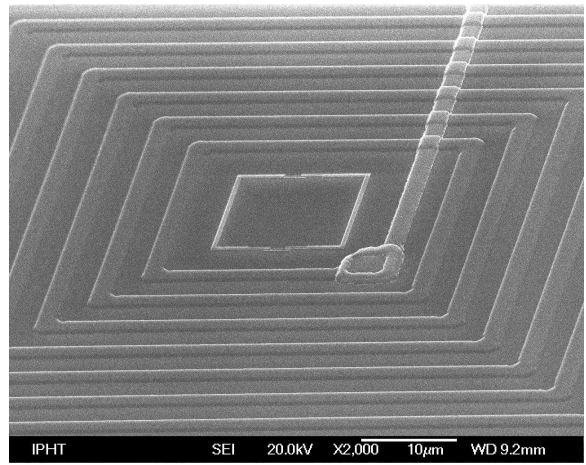}
\caption{The Al qubit inside the Nb pancake coil.}\label{fig2}
\end{figure}

The qubit was fabricated out of Al inside the tank's pickup coil
(Fig.~\ref{fig2}). We aimed for the parameters suggested in Ref.~\cite{Mooij1}. Two junctions have areas $200\times600$nm$^2$ and one is smaller, so that the critical currents have ratio $\alpha\equiv I_\mathrm{c3}/I_\mathrm{c1,2}\approx0.8$. The tank
was fabricated using a Nb thin-film pancake coil on an oxidized Si substrate. It has $\omega_\mathrm{T}/2\pi=6.284$MHz, consistent with the estimated $L_\mathrm{T}=0.2\mu$H and $C_\mathrm{T}=3$nF. The qubit has $L_\mathrm{q}=24$pH; the larger junctions have $C_\mathrm{q}=3.9$fF and $I_\mathrm{c1,2}=600$nA. The tank's quality factor $Q_\mathrm{T}\equiv\omega_\mathrm{T}/2\gamma_\mathrm{T}=1850$
corresponds to a linewidth $\sim$1.7kHz, while $M=70$pH. The noise temperature of the cold amplifier is $\sim$300mK~\cite{Oukhansky02}.

The qubit loop encloses an external magnetic
flux~$\Phi_\mathrm{e}$. Its total Josephson energy is
$E_\mathrm{J} = \sum_{i=1}^3E_{\mathrm{J}i}(\phi_i)$, where
$\phi_i$ and $E_{\mathrm{J}i}=\hbar I_{\mathrm{c}i}/2e$ are the
phase difference and Josephson energy of the $i$th junction. Due
to flux quantization and negligible~$L_\mathrm{q}$, only
$\phi_{1,2}$ are independent, with $\phi_3=-\phi_1-\phi_2-2\pi\Phi_\mathrm{e}/\Phi_0$ ($\Phi_0=h/2e$ is the flux quantum). The quantum dynamics in the resulting potential
$E_\mathrm{J}(\phi_1,\phi_2)$ is determined by the ratio of
$E_\mathrm{J}$ and the Coulomb energy $E_C \sim
e^2\!/C_\mathrm{q}\ll E_\mathrm{J}$. For suitable
parameters, at $\Phi_\mathrm{e}=\frac{1}{2}\Phi_0$ the system has
degenerate classical minima $|L\rangle$ and $|R\rangle$,
with persistent currents circulating in opposite
directions~\cite{Mooij1}. A finite $E_C$ lifts the degeneracy by
allowing tunneling between these minima, and opens a gap $\Delta$
at the anticrossing between the ground and first excited
states, $|0,1\rangle=(|L\rangle{\pm}|R\rangle)/\sqrt{2}$. Hence, the
qubit can be treated as a two-level system described
by Eq.~(\ref{eq_Ham}).

To measure $S_V$, we tuned the HF signal in resonance with the
qubit~\cite{VdW}. We only found noticeable output when $\omega_{_{\rm
HF}}/2\pi=868\pm2$MHz, in agreement with the estimated splitting
$\Delta/h\sim1$GHz. One important feature of our setup is a two
orders of magnitude mismatch between $\omega_{_{\rm HF}}$ and the
readout frequency~$\omega_\mathrm{T}$. Together with the high
$Q_\mathrm{T}$, this ensures that the signal can only be due to
resonant transitions in the qubit itself. This was verified by
measuring $S_V$ also when biasing the qubit away from degeneracy.
A signal exceeding the background, that is, emission of $\sim$6MHz
photons by the qubit in response to a resonant HF field in
agreement with Eq.~(\ref{S}), was only detected when the qubit states
were almost [cf.\ below Eq.~(\ref{Gtot})] degenerate. The measurements
were carried out at $T=10$mK. No effect of radiation was observed
above 40mK (with 40mK$/hk_\mathrm{B}\approx830$MHz, i.e.\ close
to~$\Delta/h$).

\begin{figure}
\includegraphics[width=3in]{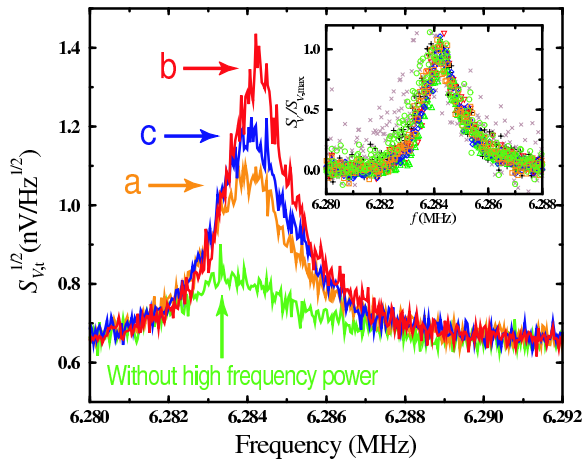}
\caption{The spectral amplitude of the tank voltage for HF powers
$P_a<P_b<P_c$ at $868$MHz, detected using the setup of
Fig.~\ref{fig1}. The bottom curve corresponds to the background
noise without an HF signal. The inset shows normalized voltage
spectra for seven values of HF power, with background
subtracted. The shape of the resonance, being determined by the
tank circuit, is essentially the same in each case. Remaining tiny variations visible in the main figure are due to the irradiated qubit modifying the tank's inductance and hence its central frequency, and in principle similarly for dissipation in the qubit increasing the tank's linewidth~\cite{footnote}; these are inconsequential for our analysis.}\label{fig3}
\end{figure}

We plotted $S_{V,\mathrm{t}}(\omega)$ for different HF powers $P$ in Fig.~\ref{fig3}. As $P$ is increased, $\omega_\mathrm{R}$ grows and passes~$\omega_\mathrm{T}$, leading to a non-monotonic dependence of the maximum signal on~$P$ in agreement with the above picture. This, and the sharp dependence on the tuning of $\omega_{_{\rm HF}}$ to the qubit frequency, confirm that the effect is due to Rabi oscillations. The inset shows that the shape is given by the second line of Eq.~(\ref{S}) for all curves.

\begin{figure}
\includegraphics[width=3.25in]{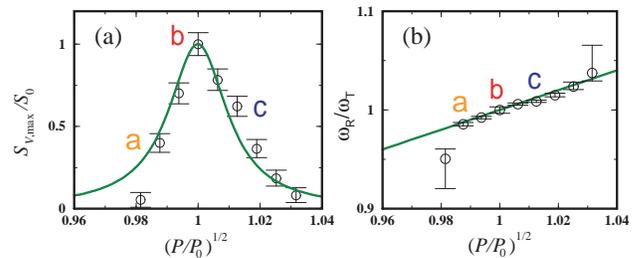}
\caption{(a)~Comparing the data to the theoretical Lorentzian. The
fitting parameter is $g\approx0.02$. Letters in the picture
correspond to those in Fig.~\ref{fig3}. (b)~The Rabi frequency
extracted from~(a) vs the applied HF amplitude. The straight line
is the predicted dependence
$\omega_\mathrm{R}/\omega_\mathrm{T}=\sqrt{P/P_0}$. The good
agreement provides strong evidence for Rabi oscillations.}\label{fig4}
\end{figure}

For a quantitative comparison between theory and experiment, we subtracted the background without an HF signal from the observed $S_{V,\mathrm{t}}$, yielding
$S_V(\omega)=S_{V,\mathrm{t}}(\omega)-S_\mathrm{b}(\omega)$~\cite{Sb}. Subsequently, we extracted the peak values vs HF amplitude, $S_{V,\mathrm{max}}(\sqrt{P/P_0})=\max_{\omega}S_V(\omega)\approx
S_V(\omega_\mathrm{T})$, where $P_0$ is the power causing the maximum response; see Fig.~\ref{fig4}a. In the same figure, we plot the theoretical curve for $S_{V,\mathrm{max}}$ normalized to its maximum~$S_0$,
\begin{equation}\label{S_norm}
 \frac{S_{V,\mathrm{max}}(w)}{S_{0}}  = \frac{w^2 g^2}{(w^2{-}1)^2 + g^2}
  \approx \frac{(g/2)^2}{(w{-}1)^2 + (g/2)^2}\;;
\end{equation}
$w\equiv\omega_\mathrm{R}/\omega_\mathrm{T}$ ($=\sqrt{P/P_0}$ theoretically) and $g=\Gamma/\omega_\mathrm{T}$. The best fit is found for
$\Gamma\approx0.02\omega_\mathrm{T}\sim8\cdot\nobreak10^5\mathrm{s}^{-1}$~\cite{footnote2}.
Thus, the life-time of the Rabi oscillations is at least
$\tau_\mathrm{Rabi}=2/\Gamma\approx2.5\mu$s~\cite{tau}, leading to an effective quality factor $Q_\mathrm{Rabi}=\Delta/(\hbar\Gamma)\sim7000$. These values substantially exceed those obtained recently for a modified 3JJ qubit ($\tau_\mathrm{Rabi}\sim150$ns)~\cite{NakamuraDelft}, which is not surprising. In our setup the qubit is read out not with a dissipative DC-SQUID, but with a high-quality resonant tank. The latter is weakly coupled to the qubit ($k^2\sim10^{-3}$), suppressing the noise leakage to it.

Finally, we can restore $\omega_\mathrm{R}(\sqrt{P})$. Assuming the spectrum to be described by Eq.~(\ref{S_norm}) with $\Gamma=0.02\omega_\mathrm{T}$, we correlate the experimental and theoretical points giving the same normalized value:
$[S_{V,\mathrm{max}}(\omega_\mathrm{R}/\omega_\mathrm{T})/
S_0]_{\rm theor} =[S_{V,\mathrm{max}}(\sqrt{P/P_0})/ S_0]_{\rm exp}$. Figure~\ref{fig4}b confirms the linear dependence $\omega_\mathrm{R}/\omega_\mathrm{T}=\sqrt{P/P_0}$ [Eq.~(\ref{eq_Rabi_freq})], given by the line.

In conclusion, we have observed Rabi oscillations in a three-junction flux qubit coupled to a resonant tank circuit. The oscillation frequency depends linearly on the applied HF amplitude, as predicted. The method of Rabi spectroscopy does not require precise pulse timing. A lower bound on the coherence time is 2.5$\mu$s. The qubit's coherence survives its coupling to the external tank circuit, to the extent that $\tau_\mathrm{Rabi}$ is $\sim$16 times the Rabi period. All main features have been reproduced in a second sample with $\omega_{_\mathrm{HF}}/2\pi=3.665$GHz and $\omega_\mathrm{T}/2\pi=27.5$MHz (data not shown), though with a somewhat lower $Q_\mathrm{Rabi}$. Thus, high-quality superconducting tanks provide a straightforward and sensitive method for qubit characterization; their use as coupling elements between qubits is actively being researched~\cite{couple}.

\begin{acknowledgments}
We thank D.~Born and H.~M\"uhlig for technical assistance; U.~H\"ubner, T.~May, and
I.~Zhilyaev for sample fabrication; and I.~Chiorescu, Ya.S. Greenberg, W.~Hardy, J.P.
Hilton, H.E. Hoenig, Y.~Imry, W.~Krech, Y.~Nakamura, G.~Rose, A.~Shnirman, P.C.E. Stamp, M.F.H. Steininger, and W.G. Unruh for fruitful discussions.
\end{acknowledgments}


\begin{thebibliography}{99}
\bibitem{Sb} Without an HF signal, the qubit's influence at $\omega_\mathrm{T}$ is negligible. Thus, the ``dark'' trace in Fig.~\ref{fig3} is a quantitative measure of~$S_\mathrm{b}$.
\bibitem{footnote2} Since the tank damping is $\gamma_\mathrm{T} \sim 3000\mathrm{ s}^{-1}$, this justifies our approximation $\Gamma \gg \gamma_\mathrm{T}$.
\bibitem{tau} This may in fact exceed the intrinsic ``free-induction'' coherence time $\tau_\phi$, as already evidenced by the factor 2 in~$\tau_\mathrm{Rabi}$. Cf.\ the concluding remarks in~\cite{NakamuraDelft}.
\end{thebibliography}
\end{document}